# Generic Extensional Framework for the Memristive Systems

Zhiwei Li, Haijun Liu, and Xin Xu

*Abstract*—Considering the over-requirement of continuity and "well-defined" confined by the canonical framework model for the memristive systems, this letter proposes a more generic extensional framework to accommodate those discontinuity or not "well-defined" situations. In particular, two special subtypes are pointed out to fit the hitherto memristor device models.

## I. Introduction

SINCE HP Labs claimed their successful invention of memristor in about half decade ago, extensive attention has been draw back to the memristor as the next generation of non-volatile memory. Albeit its first postulation by Chua dates back to 1970s, both theoretical development and physical exploration of nano-devices are yet hot issues. And a myriad of hitherto research prefer the notion of general two-terminal memristive system to that of ideal memristor and primarily concentrate on the canonical family of time-invariant memory system:.

$$y(t) = h(\mathbf{x},u) = g(\mathbf{x},u)u(t), \quad d\mathbf{x}/dt = \mathbf{f}(\mathbf{x},u). \quad (1)$$

where $\mathbf{x}$ denotes a vector with $n$ internal state variables of system, $y$ and $u$ represent the output and input of system respectively. The symbol $\mathbf{f}$ is a continuous $n$-dimensional vector function, while the continuous scalar function $h$ can be explicitly factored as the product of n-dimensional scalar function $g$ and system input $u$.

It is notable that Chua and Kang [1] originally defined the property of continuity acquired by $g$ and $\mathbf{f}$ that results in the zero-crossing distinctive feature. Later in 2011, Chua [2] emphasized $g(\mathbf{x},0) \neq \infty$ to ensure the $y$-$u$ hysteresis loop passing through the origin. Moreover, a concept of "well-defined", which means $g \neq 0$ and $g \neq \infty$, was introduced by Pershin and Ventra [3] to guarantee that salient fingerprint. In contrast to requirement of continuity for $g$, several discontinuous dynamical systems were demonstrated by Mouttet [9]. Although those examples exhibiting a pinched hysteresis loop in the $y$-$u$ plane merely suit for the input signal with certain sinusoidal form which was pointed out and debated by Kim etc. [10], yet an absorbing question is raised that whether the canonical framework is a must for the memristive system? This letter provides an extensional framework to give a positive answer to this question. Such framework model allows the discontinuity caused by the input $u$ not the state variables $\mathbf{x}$, which as result avoids the constraint on the input form.

## II. Proposed Framework

The proposed framework model for one-port memristive system is specified by the following equations:

$$y(t) = h(\mathbf{x},u) = <\mathbf{g}(\mathbf{x},u), \mathbf{m}^T(u)>, \quad d\mathbf{x}/dt = \mathbf{f}(\mathbf{x},u). \quad (2)$$

where $\mathbf{x}$, $y$ and $u$ still hold the same meanings in (1). The major change consists in that the continuous scalar function $h$ shall be explicitly factored as the inner product of two continuous $p$-dimensional vector functions, which is similar to variable separation method. The function $\mathbf{g}: \Re^n \times \Re \to \Re^p$ concerns both the inner state variables and the input and vector function $\mathbf{m}: \Re \to \Re^p$ merely concerns the input excitation meanwhile holding the zero-crossing property, i.e. $\mathbf{m}(0) = \mathbf{0}$. In special case when the dimension $p=1$ and $\mathbf{m}(u)=u$, (2) reduces to the canonical framework. To make it more sensible, we enforce the transformation from (2) to (1):

$$\begin{aligned}y(t) = h(\mathbf{x},u) &= <\mathbf{g}(\mathbf{x},u), \mathbf{m}^T(u)> = <\mathbf{g}(\mathbf{x},u), \mathbf{m}^T(u)/u > u \\ &= <\mathbf{g}(\mathbf{x},u), \mathbf{m}_*^T(u)> u = g_*(\mathbf{x},u)u.\end{aligned} \quad (3)$$

It follows from (3) that the new response function $g_*(\mathbf{x},u)$ exhibits no promise of the continuity and "well-defined" but it's still able to capture the characteristic of pinched hysteresis loop as a result of the continuity for output function $h$ and the zero-crossing property for $\mathbf{m}$.

The non-symmetric phenomenon of $y$-$u$ Lissajous figure could occur according to the different choice of vector function $\mathbf{m}$. Apart from the influence of function $\mathbf{f}$ which is termed "memristor morphing function" [2], function $\mathbf{m}$ additionally plays a key role in the capability of deforming and morphing the odd-symmetric pinched hysteresis loops into other non-symmetrical shapes owing to the contribution of input.

In contradistinction to the traditional canonic framework, two special degenerate cases of extensional framework with one single inner state variable would be pointed out:

$$\begin{aligned}y(t) = h(x,u) &= \mathbf{g}(x,u)\mathbf{m}^T(u) = g_*(x,u)u \\ g_*(x,u) &= \mathbf{g}(x,u)\bigl(\mathbf{m}^T(u)/u\bigr)\end{aligned} \quad (4)$$

Manuscript received May 2, 2012. This work was supported in part by the National Nature Science Fund Committee under Grant 61171017 and the Graduate Student Innovation Fund of NUDT under Grant S120402.

Z.W. Li, H.J. Liu, and X. Xu are with the College of Electronic Science and Engineering, National University of Defense Technology, Changsha, CO 410073 China (e-mail: lzw89523@gmail.com; liuhaijun@nudt.edu.cn; xuxin@nudt.edu.cn).

where $m(0)=0$, $h(\mathbf{x},0)=0$ and the vital issue falls on the properties of $\mathbf{m}(u)$, and

$$\mathbf{m}(u) = (m_1(u), m_2(u), \ldots, m_p(u)) \quad (5)$$

If $u=0$ belongs to the discontinuity point of the first kind of function $m_i(u)/u$, $i=1,2,\ldots,p$, we define this memristive system with type I extensional framework which disobeys the original canonical framework in [1]. As defined, the type I extensional framework contains two subclasses. The first one is $\lim_{u\to 0} m_i(u)/u$ converges to a constant while $m_i(u)/u$ is meaningless at $u=0$. One such framework example has been demonstrated by J. J. Yang etc. [4] to model the memristor behavior of coupled electron-ion dynamics existed in metal/oxide/metal nano-switches:

$$I(t) = x^n \beta_1 \sinh(\alpha_1 V(t)) + \beta_2 (\exp(\alpha_2 V(t)) - 1) \quad (6)$$
$$dx/dt = V(t)$$

where $\alpha_i$ and $\beta_i$ are fitting constants, the integer $n$ is a free parameter. By rearranging terms to obtain the form of (3):

$$I(t) = \begin{bmatrix} x^n \beta_1, \beta_2 \end{bmatrix} \begin{bmatrix} \sinh(\alpha_1 V) \\ \exp(\alpha_2 V) - 1 \end{bmatrix}$$
$$= \left( \begin{bmatrix} x^n \beta_1, \beta_2 \end{bmatrix} \begin{bmatrix} \sinh(\alpha_1 V)/V \\ (\exp(\alpha_2 V)-1)/V \end{bmatrix} \right) V \quad (7)$$

We observe from (7) that in the view of canonical framework, such a memristive nano-device contains a serial breakpoint at $V=0$ called removable discontinuity (maybe more properly termed singularity) in mathematics, which is in fact not a canonical memristive system but rather a memristive system with type I extensional framework. The model results in Fig. 1 show the non-symmetrical manner caused by $\exp(\alpha_2 V)-1$ with non-symmetrical property.

The other situation occurs when $\lim_{u\to 0} m_i(u)/u$ converges to more than just one value. Since there would have to be two different limits depending on the state variable which is obvious graphically from the conventional pinched hysteresis curve which has two different slopes at $u=0$ corresponding to two different resistance states. One such example has been proposed in [8] to model synapse for the neuromorphic systems:

$$I(t) = \begin{cases} a_1 x(t) \sinh(b_1 V(t)), & V(t) \geq 0 \\ a_2 x(t) \sinh(b_2 V(t)), & V(t) < 0 \end{cases}$$
$$\frac{dx}{dt} = \begin{cases} c_1 \sinh(d_1 V(t)), & V(t) \geq 0 \\ c_2 \sinh(d_2 V(t)), & V(t) < 0 \end{cases} \quad (8)$$

where $a_i$, $b_i$, $c_i$ and $d_i$ are all positive adjustable constants. By rearranging terms to obtain the $g_*$ in (4):

$$g_*(x,V) = \begin{cases} a_1 x(t) \sinh(b_1 V(t))/V, & V(t) > 0 \\ a_2 x(t) \sinh(b_2 V(t))/V, & V(t) < 0 \end{cases} \quad (9)$$

Commonly when $a_1 b_1 \neq a_2 b_2$, as the simulation results shown in Fig. 2, we observe that the memristive synapse model contains a serial breakpoint at $V=0$ called jump discontinuity (the limits from the left and right both exist but are not equal to each other) in mathematics. In addition, a not "well-defined" situation $g_*(x,V)|_{x=0} = 0$ may occur in this model. Other similar device models of type I could be found in [2], [6] and [7].

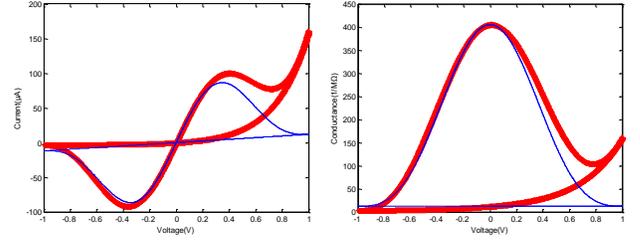

Fig. 1. Simulated *V-I* curves and hysteresis loops of conductance associated with voltage. In our Matlab simulation, $\alpha_1 = 2$, $\alpha_2 = 4$, $\beta_1 = 3e-9$, $\beta_2 = 3e-6$, $n=16$ and the applied input $V=\sin t$. The red dotted curves are obtained according to (6) and the blue solid cures are obtained by replacing $\sinh(\alpha_1 V)$ and $\exp(\alpha_2 V)-1$ with $\alpha_1 V$ and $\alpha_2 V$ respectively.

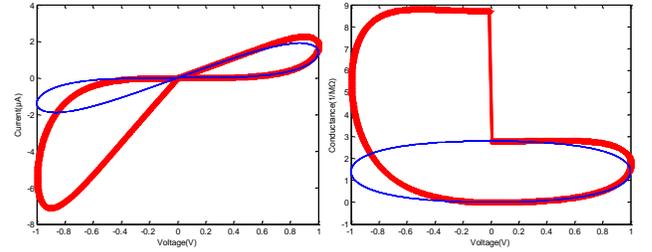

Fig. 2. Simulated *V-I* curves and hysteresis loops of conductance associated with voltage. In our Matlab simulation, $b_1 = b_2 = 1.2$, $c_1 = c_2 = 6e-4$, $d_1 = d_2 = 2$ and the applied input $V=\sin t$. The blue solid curves are obtained by setting $a_1 = 4e-3$ and $a_2 = 1.25e-2$, while the red dotted curves are obtained by setting $a_1 = a_2 = 4e-3$ and replacing $\sinh(b_i V(t))$ with $b_i V(t)$.

Another type is generated when the zero order of $m_i(u)$ is lower than that of $u$, i.e. there exists $\lim_{u\to 0} m_i(u)/u = \infty$, $i=1,2,\ldots,p$. we call such being not "well-defined" memristive system with type II extensional framework which is outside of the constraint $g(\mathbf{x},u) \neq \infty$ underscored by Chua in [3]. This situation seems to be discussed less, however, here let's consider a memristive system constructed by the ideal HP's memristor [5] and the diode clipper as shown in Fig. 3.

$$V = \left( R_{ON} \frac{x(t)}{D} + R_{OFF} \left(1 - \frac{x(t)}{D}\right) \right) I(t) + R_d(I) I$$
$$= \left[ R_{ON} \frac{x(t)}{D} + R_{OFF} \left(1 - \frac{x(t)}{D}\right), 1 \right] \begin{bmatrix} I \\ R_d(I) I \end{bmatrix} \quad (10)$$
$$\dot{x} = \mu_V \frac{R_{ON}}{D} I(t)$$

where $R_d$ represents the resistance of diode determined by the current value. As the numerical simulation results shown in Fig. 4, when $I > 0$, the $R_d(I)I$ almost equals a constant $V_T$ which often is referred as a positive threshold voltage of the diode to switch the on/off state of the diode. Otherwise $I \approx 0$, $R_d(I) \approx \infty$ which indicates a memristive system with type II extensional framework instead of a canonical one.

III. CONCLUDING REMARKS

The proposed extensional framework is straightforward manifestation of the existence of non-canonical memristive systems, which further shows both the inner state and the input contribute to the resistance for a memristive system. It should be more suitable for the extensional framework than the canonical one to model many memristive devices and systems. More detailed analysis on the characteristic will be reported in a forthcoming paper.

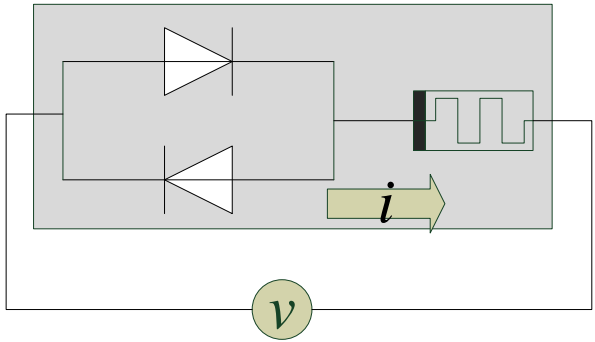

Fig. 3. The one port memristive system constructed by the memristor and the diode clipper

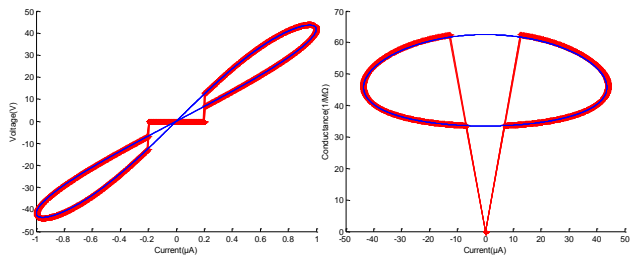

Fig. 4. Simulated *V-I* curves and hysteresis loops of conductance associated with current. In our Matlab simulation, $R_{ON}=100\,\Omega$, $R_{OFF}=16\,\mathrm{k}\Omega$, $D=1e-8$, $\mu_v=1e-14$ and the applied input $V=\sin t$. The blue solid curves are obtained by setting $V_T=0.2$, while the red dotted curves are obtained by setting $V_T=0$, namely neglecting the clipper effect.

REFERENCES

[1] L. O. Chua, and S. M. Kang, "Memristive devices and systems," *Proc. IEEE*, vol. 64, no. 2, Feb. 1976.
[2] L. O. Chua, "Resistance switching memories are memristors," *J. Appl. Phys. A*, vol 102, no. 4, Jan 2011.
[3] Y. V. Pershin and M. D. Ventra, "Memory effects in complex materials and nanoscale systems," *Adv. Phys.*, vol. 60, pp. 145-227, 2011.
[4] J. J. Yang, M. D. Pickett, X. Li, D. A. A. Ohlberg, D. R. Stewart, and R. S. Williams, "Memristive switching mechanism for metal/oxide/metal nanodevices," *Nat. Nanotechnol.*, vol.3, no.7, pp. 429–433, Jul. 2008.
[5] D. B. Strukov, G. S. Snider, D. R. Stewart, and R. S. Williams, "The missing memristor found," *Nature*, vol. 453, no. 7191, pp. 80–83, 2008.
[6] C. Yakopcic, T. M. Taha, G. Subramanyam, R. E. Pino, and S. Rogers "A memristor device model," *IEEE Electron Device Lett.*, vol. 32, no. 10, pp 1436-1438, Oct. 2011.
[7] J. G. Simmons, "Generalized formula for the electric tunnel effect be-tween similar electrodes separated by a thin insulating film," *J. Appl. Phys.*, vol. 34, no.6, p.1793, Jun.1963.
[8] M. Laiho, E. Lehtonen, A. Russel, and P. Dudek, "Memristive synapses are becoming a reality," *The Neuromorphic Engineer*. College Park, MD: Inst. Neuromorphic Eng., 2010.
[9] B. Mouttet, "Memresistors and non-memristive zero-crossing hysteresis curves," arXiv:1201.2626v5, Mar, 2012.
[10] H. Kim, M. Sah, and S. P. Adhikari, "Pinched Hysteresis Loops is the Fingerprint of Memristive Devices", arXiv:1202.2437, Feb, 2012.